\documentclass[12pt]{article}
\usepackage{amsmath,latexsym,epsfig,amssymb}

\input eqalign.sty     

\def\ppall{\mathaccent23p}
\def\ppall{\mathaccent23p}

\def\qpall{\mathaccent23q}

\newcommand{\D}{\mathcal{D}}

\newcommand{\dds}{\stackrel{\leftrightarrow}{D}}

\newcommand\be{\begin{equation}}
\newcommand\ee{\end{equation}}
\newcommand\bea{\begin{eqnarray}}
\newcommand\eea{\end{eqnarray}}

\def\ma[#1,#2,#3,#4]  {{\left( \matrix{ #1  & #2 \cr
                                        #3  & #4 \cr } \right)}}

\begin{document}

\title{
{\vspace{-1cm} \normalsize
\hfill \parbox{40mm}{CERN/TH-98-265}}\\
Non-perturbative running of the average momentum of non-singlet parton densities }
\author{
Marco Guagnelli$^1$,  
Karl Jansen$^{2,}$\footnote{Heisenberg Foundation Fellow}  
$\;$ and Roberto Petronzio$^{1,}$$^2$  
\\
{\footnotesize $^1$ Dipartimento di Fisica, Universit\`a di Roma {\em Tor Vergata} }\\
{\footnotesize and INFN, Sezione di Roma II} \\
{\footnotesize Via della Ricerca Scientifica 1, 00133 Rome, Italy } \\
{\footnotesize 
$^2$ CERN, 1211 Geneva 23, Switzerland  
}
}
\maketitle

\begin{abstract}
We determine 
non-perturbatively the anomalous dimensions of the second  moment of non-singlet parton densities
from a continuum extrapolation of results computed in quenched lattice simulations 
at different lattice spacings.
We use a Schr\"odinger functional scheme for the definition of the renormalization constant of
the relevant twist-2 operator. In the region of renormalized couplings explored, we obtain a good
description of our data in terms of a three-loop expression for the anomalous dimensions. 
The calculation can be used for exploring values of the coupling where a perturbative expansion of
the anomalous dimensions is not valid a priori. Moreover,  our results provide the non-perturbative 
renormalization constant that connects hadron matrix elements on the lattice, renormalized 
at a low scale, with the experimental results, renormalized at much higher energy scales.
\end{abstract}

\section{Introduction}

The accurate knowledge of hadron parton densities is an essential ingredient for the experimental
tests of QCD at accelerator energies. Their normalization is usually obtained from a fit to a set
of reference experiments and used for predicting the behaviour of hard hadron processes in different
energy regimes. The energy evolution and the relative normalization of the hard processes are predictable
within renormalization-group-improved perturbation theory, while the calculation of the absolute normalization 
needs non-perturbative methods.
These methods may provide not only a check of the non-perturbative aspects
of QCD but also some information that could help in fixing the values of
 parton densities at values of the Bjorken $x$ larger than $0.5$ where the experimental
information, especially for the gluon distribution, is scarce \cite{phenomeno}. 
Besides, they could also help in
establishing the domain of validity of a perturbative evolution that appears in some cases
to be compatible down to scales where one would expect higher orders and power corrections
to inelastic processes to take over.  A study of these
non-perturbative effects in moments of hadronic structure functions
can be done within the lattice approach to QCD 
\cite{martinelli,schierholz1,schierholz2,schierholz3}. 

In this paper we determine 
the non-perturbative anomalous dimensions 
of the second moment of non-singlet parton densities.
The anomalous dimensions are extracted 
from a continuum extrapolation of results obtained in quenched lattice simulations at different 
lattice spacings. 
The definition of the renormalization constant and of the matrix element of the relevant
twist-2 operator is done within the finite-volume Schr\"odinger functional (SF) scheme
\cite{schrfunc,sint,paper3,letter}.
Non-perturbative renormalization constants within this scheme have  been obtained
by the ALPHA collaboration for local fermion bilinears, and
in particular for the scale-dependent renormalization constant 
 $Z_{\rm P}$ entering the computation of the running quark mass \cite{quarkmass}.
Here we follow ref.~\cite{roberto_pert}, where the perturbative calculations needed to relate
the SF scheme to more common schemes, such as the modified minimal subtraction scheme, have
been carried out and some general considerations on the onset of lattice artefacts have been presented.

In the region of couplings that we explore,
we find a good agreement between our data and a three-loop approximation for the anomalous dimensions
in the SF scheme.
As a consistency check, we successfully compare the running of our renormalization constant 
with the one expected from a fit to the anomalous dimensions.

In section 2 we recall some basic facts about the Schr\"odinger functional scheme,
we define the matrix element of the twist-2 operator and we outline the general
strategy of the calculation. Section 3 contains the details of our numerical simulations and the
results for the continuum extrapolated values of the ratio of renormalization constants,
i.e. the step scaling functions, calculated
at renormalization scales differing by a factor 2.
In section 4 we extract from these numerical results a fit to the renormalized coupling constant
dependence of the anomalous dimensions of the operator. 
Section 5 is devoted to the conclusions.

\section{The Schr\"odinger functional scheme and the twist-2 operator}

The Schr\"odinger functional has been discussed extensively in the
literature (see \cite{rainer_lecture,martin_lecture} for reviews); 
it represents the (quantum) amplitude for the time evolution 
of a field configuration between 
two predetermined classical states at some time boundaries. 
It takes the form of a standard functional integral with fixed boundary 
conditions.
It has been shown that its renormalizability properties are the same as those 
of the theory with infinite time extension, 
modulo the possible presence of a finite 
number of boundary counterterms \cite{sint}.
In QCD with fermions, it can be written as:

\begin{equation}
\mathcal{Z}[C^{'},\bar{\rho}',\rho';C,\bar{\rho},\rho] = \int \D[U] \D[\psi]
\D[\bar{\psi}] \rm{e}^{-S[U,\bar{\psi},\psi]}\; ,
\label{eq:sf}
\end{equation}

\noindent where $C^{'},C$ and $\bar{\rho}',\rho',\bar{\rho},\rho$ are the boundary 
values of the gauge and fermion fields respectively.
In the following discussion, the classical boundary gauge field will be set 
to zero.
According to refs.~\cite{letter,paper1}, expectations values may involve the 
response $\zeta $ to a variation of the classical Fermi field configurations on the 
boundaries:

\begin{equation}
\zeta({\bf{x}}) = \frac{\delta}{\delta\bar{\rho}({\bf{x}})}, \qquad
\bar{\zeta}({\bf{x}}) = -\frac{\delta}{\delta\rho({\bf{x}})}
\end{equation}

\begin{equation}
\zeta'({\bf{x}}) = \frac{\delta}{\delta\bar{\rho}'({\bf{x}})}, \qquad
\bar{\zeta}'({\bf{x}}) = -\frac{\delta}{\delta\rho'({\bf{x}})}\; .
\end{equation}

In the continuum, moments of non-singlet structure functions are related, 
through the operator product expansion, to hadronic matrix elements of local 
twist-2 operators of the form:

\begin{equation}
{\cal{O}}_{\mu_{1}\ldots \mu_{n}}^{qNS} = 
\bigl( \frac{i}{2}\bigl)^{n-1} {\bar{\psi}}(x)\gamma_{\{\mu_{1}} 
\dds_{\mu_{2}}\cdots \dds_{\mu_{n}\}} 
\frac{\lambda^f}{2} \psi(x)\ +\ \mbox{trace terms}\; ,
\label{eq:twist two_continuum}
\end{equation}

\noindent where $\dds_{\mu}$ is the covariant derivative, 
$\{$``indices''$\}$ means symmetrization. 

The twist is defined as the difference between the engineering dimensions 
of the operator and its angular momentum. 
All listed operators belong to irreducible representations 
of the Lorentz group. 

On the lattice, the discretization of the covariant derivative can be done in 
a standard way:

\begin{equation}
\meqalign{
\bigtriangledown_{\mu} \psi(x) = 
\frac{1}{a}[U_\mu(x) \psi(x+a\hat{\mu}) - \psi(x)] \cr
\bigtriangledown^{\dagger}_{\mu} \psi(x) = 
\frac{1}{a}[\psi(x)-U_\mu(x-a\hat{\mu})^{-1} 
\psi(x-a\hat{\mu})]\; .
\label{eq:D_discrete}
}
\end{equation}

In this paper, we concentrate on the calculation of the 
second moment to which we associate the following irreducible operator that
is multiplicatively renormalizable:

\begin{equation}\meqalign{
O^q_{12} = \frac{i}{2}\bar\psi \gamma_{\{1} \dds_{2\}}\frac{\lambda^f}{2}\psi.
\label{eq:lat_operators}
}\end{equation}

We define the  SF matrix element of the second moment of the non-singlet parton
densities by the 
observable:

\begin{equation}\label{corrfunc}\meqalign{
f_2(x_0) \equiv f_{0_{12}}(x_0) = -a^6\sum_{\bf{y},\bf{z}} \rm{e}^
{i\bf{p}(\bf{y}-\bf{z})}
\langle \frac{1}{4} \bar\psi(x) \gamma_{[1} 
\dds_{2]}\frac{1}{2} \tau^3 \psi(x) 
\bar\zeta({\bf{y}}) \Gamma \frac{1}{2} \tau^3 \zeta({\bf{z}})\rangle
}
\end{equation}

\noindent where the contraction of the classical fields is non-vanishing if the matrix 
$\Gamma $ satisfies: $\Gamma P_{-(+)} = P_{+(-)}$, 
where $P_{-(+)}=\frac{1}{2}(1\pm\gamma_0)$ and ${\bf p}$ is the 
momentum of the classical field sitting on the boundary. The observable can be seen as the
operator matrix elements between the vacuum and ``$\rho$''-like classical states sitting
at the $T=0$ boundary.

We take the limit of massless quarks, which in the numerical simulations
can be monitored  via axial Ward identities. In the
SF framework it is possible to work at zero physical quark mass
because a natural infrared cutoff to the Dirac operator eigenmodes is
provided by the time extent of the lattice. 

The matrix element of the operator for the second moment involves 
two directions:
 one of them is given by the contraction matrix $\Gamma $,
i.e. from the polarization of the vector classical state:

\begin{equation}
\Gamma = \gamma_2,  
\label{eq:gamma}
\end{equation}

\noindent and the other one from the momentum $p$ of the classical Fermi 
field at the boundary.
For the tree level, this choice gives a non-vanishing matrix element 
in the massless quark limit, where we evaluate our correlations. 
The tree-level correlation can be easily calculated
and reads:

\begin{equation}
f^{(0)}_2(x_0) = \frac{i\ppall_1 N}{R(p)^2} \left[ (-i\ppall_0) 
\left(M_-(p) {\rm{e}}^{-2\omega({\bf{p}})x_0} - 
M_+(p) {\rm{e}}^{-2\omega({\bf{p}})(2T-x_0)}\right) 
\right]\; ,
\label{eq:tree}
\end{equation}

\noindent where 

\begin{equation}
\hat p_\mu = (2/a) \sin (ap_\mu/2), \qquad \ppall_\mu = (1/a) \sin (ap_\mu)\; ,
\end{equation}

\begin{equation}
M(p) = m + \frac{1}{2}a \hat{p}^2, \qquad M_{\pm} = M(p) \pm i \ppall_0\; ,
\end{equation}

\begin{equation}
-i\sin(ap_0/2)\equiv \textrm{sinh } \left[\frac{a}{2}\omega({\bf{q}})\right] = 
\frac{a}{2} \left\{ \frac{ {\bf{\qpall}} ^2 + 
(m +\frac{1}{2} a\hat{{\bf{q}}}^2)^2}
{1+a (m +\frac{1}{2} a\hat{{\bf{q}}}^2)} \right\}^{\frac{1}{2}}\; ,
\end{equation}

\begin{equation}
R(p) = M(p) 
\left\{1-\textrm{e}^{-2\omega({\bf{p}})T}\right\} -
i\ppall_0 \left\{1+\textrm{e}^{-2\omega({\bf{p}})T}\right\}.
\end{equation}

We have chosen as convention 
$\Gamma = \gamma_{2}$ and ${\mathbf p} = (p_1,0,0)$.
Furthermore, we will always work on hypercubic lattices
with $T=L$.

We may define an unnormalized observable 
as the ratio of the correlation function $f_2(x_0)$ 
divided by its tree-level expression   
\begin{equation}
\bar{Z}(L) = \frac{f_2(\eta L)}{f_2^{(0)}(\eta L) }\; ,
\end{equation}
with $\eta<1$ and typically $\eta=1/2$ or $\eta=1/4$ (see below).  
The normalized observable is defined by removing the 
renormalization constant of the classical boundary sources $\zeta$.      
Following refs.~\cite{paper4,peterstefan}, this is represented by
the quantity called $f_1$, also normalized by its tree-level 
expression.
We thus arrive at our definition of the renormalization constant for the 
twist-2 non-singlet operator:
\begin{equation}\label{finalZ}
Z(L) = \frac{\bar{Z}(L)}{\sqrt{f_1(L)}}\; .
\end{equation} 

We denote by $L$ the physical lattice size and by $a$ the lattice cut-off. 
The observables on the lattice depend
upon the ratio $L/a$, the value of the momentum $ap_1$,
and the point $x_0/a$ (or equivalently the value of $\eta$)
of the insertion of the operator in time.
We note that, in fact, time translation
symmetry is broken by the fixed boundary conditions. 
In the following, both $p$ and $x_0$ will be taken proportional to $1/L$ and $L$,
respectively, so that the only external scale in the problem is $L$. 
In the expressions for $Z$, $\bar{Z}$ and $f_1$ the lattice spacing units 
are understood, and we will make them explicit only when needed.

The operator needs renormalization to be finite in the continuum limit:
we define the renormalization constant such that the operator 
matrix element is equal to its tree-level value at $\mu = 1/L$.
At one-loop, we obtain:

\begin{equation}\meqalign{
O^{R}(\mu) = (1 - g^2 Z^{(1)}(1/a\mu))O^{\rm bare}(a) \cr
= (1 - g^2 Z^{(1)}(1/a\mu))(1 + g^2 Z^{(1)}(L/a)) O^{\rm tree}\; ,
\label{eq:ren_condition}
}\end{equation}

\noindent where $O^{\rm tree}$ is the result at zero coupling.
Beyond perturbation theory, the renormalized operator is defined by:

\begin{equation}
O^{R}(\mu) = Z(1/a\mu)^{-1}O^{\rm bare}(a/L)
\end{equation}

with $Z(L/a)$ defined by:

\begin{equation}
O^{\rm bare}(a/L) = Z(L/a) O^{\rm tree}\; ,
\end{equation}

where we have explicit the lattice cut-off dependence.
The study of the scale dependence of the renormalization constant
is then equivalent to that of the dependence upon the lattice size $L$,
provided that, as anticipated, the external variables, 
upon which the matrix element may depend, scale like the
basic length $L$ in physical units: we hence take 
\begin{equation}
p_1 =  2 \pi/L \cr
x_0 = \eta L, 
\end{equation}

\noindent with $\eta = 1/4$ or $1/2$.

Following ref.~\cite{roberto_pert}, the renormalization constant up to one-loop, $Z^{(1)}(L/a)$, 
in the continuum can be parametrized by:

\begin{equation}\label{Z_pert}
Z^{(1)}(L/a) = B_0 + C_0 \ln(L/a)\; , 
\end{equation}
\noindent with
\begin{eqnarray}
B_0 & = & 0.2635(10)\;\; [\eta = 1/4] \nonumber \\
B_0 & = & 0.2762(5)\;\;\;\;  [\eta = 1/2]\; .
\end{eqnarray}

The central goal of the calculation consists in obtaining the continuum limit of
the so-called ``step scaling function'' $\sigma_{\rm Z}$,
describing the change of $Z$ when the renormalization scale,
proportional to the physical box size $L$, is varied
by a factor of $s$, i.e.
\begin{equation}\label{Zs}
Z(sL) = \sigma_Z(\bar{g}^2(L))Z(L)\; ,
\end{equation}
at a fixed value of the running coupling $\bar{g}(L)$ renormalized at the scale $L$.
The quantity $\sigma$ expressed in terms of the renormalized coupling constant
does not depend anymore on the lattice cut-off and has a finite continuum limit. 
From the jump for a finite renormalization scale we can extract the one for an
infinitesimal variation governed by the corresponding anomalous dimension.

On the lattice, we can only obtain the continuum quantities by extrapolating the
lattice step scaling functions obtained, at fixed lattice spacing, from the 
ratio of renormalization constants computed on lattices with $N$ and $sN$ points. 
In order to extrapolate to the continuum, we perform several simulations at 
increasing values of the bare coupling constant
and decreasing values of the lattice spacing by increasing the number of lattice points $N$, 
so that the physical volume $L= Na$
remains constant. The latter condition is monitored by the value of the renormalized coupling 
constant in the SF scheme, where the renormalization scale is given by the physical
lattice extent $L$.

We have used the values of bare couplings and volumes corresponding to a fixed
$\alpha (L)$ determined by the ALPHA collaboration in their study of the running mass.
In general, reaching large values of the renormalized coupling, and therefore large physical volumes,
with a limited number of points, implies dealing with increasing lattice artefacts.

We extrapolate our lattice step scaling functions 
to the continuum limit and compare the dependence of the
continuum step scaling function on the coupling with the expression obtained from a 
perturbative expansion of the anomalous dimensions and of the $\beta$-function in the
renormalized coupling.

More explicitly, the anomalous dimensions in a scale-dependent regularization scheme (momentum
subtraction or SF scheme) are defined from the variation of the renormalization constant
to an infinitesimal variation of the scale as:

\begin{equation}
\frac{d\log(Z(\mu))}{d\log(\mu)}  = \gamma(g^2(\mu))
\end{equation}

with $g^2(\mu)$ satisfying:

\begin{equation}
\frac{dg^2(\mu)}{d\log(\mu)} = \beta(g^2(\mu))\; .
\end{equation}

In our simulations we will always choose $s=2$ in eq.~(\ref{Zs}). The ratio $Z(2L)/Z(L)$ can be  expressed as:

\begin{eqnarray} \label{sigmapert}
\frac{Z(2L)}{Z(L)} & = & \exp \left\{\int_{L}^{2L} d[\log(Z(L'))]\right\} \nonumber \\ 
& = &  
\exp\left\{ \int_{g^2(L)}^{g^2(2L)} d[g^2(L')] \frac{d\log(Z(L'))}
{d\log(L')} \cdot \frac{d\log(L')}{dg^2(L')}\right\} \nonumber \\
& = & 
\exp\left\{ \int_{g^2(L)}^{g^2(2L)} d[g^2(L')] \frac{\gamma(g^2(L'))}{\beta(g^2(L'))}\right\}\; .
\end{eqnarray}

For example, by inserting a perturbative expression to three-loop for gamma
 
\begin{equation} \label{gamma_3loop}
\gamma(g^2(\mu)) = \gamma_0 g^2(\mu) + \gamma_1 g^4(\mu) + \gamma_2 g^6(\mu),
\end{equation}
 
\noindent and for beta 
 
\begin{equation} \label{beta_3loop} 
\beta(g^2(\mu)) = \beta_0 g^4(\mu) + \beta_1 g^6(\mu) + \beta_2 g^8(\mu),
\end{equation}
 
\noindent  we get for the finite scale jump:
 
\begin{equation}\label{sigma_pert}
\log(Z(2L)/Z(L)) = F(g^2(2L)) - F(g^2(L)), 
\end{equation}
 
\noindent where 
 
 
\begin{eqnarray}
F(x) & = & \frac{\gamma_0}{2\beta_0 }\left( \log\left(\frac{x^2}{1 
            + \beta_1/\beta_0 x + \beta_2/\beta_0 x^2}\right)
- \beta_1/\beta_0 I(x)\right) 
 +  \frac{\gamma_1}{\beta_0 } I(x) \\
& + & \frac{\gamma_2}{2\beta_2 } ( \log(1 + \beta_1/\beta_0 x + \beta_2/\beta_0 x^2) - \beta_1/\beta_0 I(x))
\end{eqnarray}
 
\noindent with
 
\begin{eqnarray}
I(x) & =&  (2/\sqrt{\Delta}) {\rm arctg}\left(\frac{\beta_1/\beta_0 x + 2\beta_2/\beta_0 x}{\sqrt{\Delta}}\right), \\
\Delta & = & 4 \beta_2/\beta_0 - (\beta_1/\beta_0)^2\; ,
\end{eqnarray}
 
for $\Delta > 0$ as in our case.

\section{The step scaling functions}

In this section we present our numerical results for the non-perturbative
evaluation of the step scaling function. 
We used normal Wilson fermions without any
improvement and worked on even-sized lattices ranging from
$6^4$ to $32^4$. We employed SSOR \cite{ssor} preconditioning and
a BiCGstab solver \cite{bicg} for all necessary matrix inversions. 
The gauge fields were generated with a hybrid of heatbath
and over-relaxation updates. We normally performed 12 to 16
over-relaxation steps per heatbath update and 20 to 50
iterations in between measurements for the $16^4$ and the $32^4$
lattices, respectively. All errors quoted below are computed
using a jack-knife method. We explicitly checked by combining
the jack-knife method with a binning procedure, that there exist no noticeable
autocorrelation times for our observables.
The statistics of our data are 300 measurements for the $24^4$ and
the $32^4$ lattices and reach 500 measurements for the smaller
lattices. 
To complete the specification of the numerical simulations we performed,
we finally give some Schr\"odinger functional specific parameters
\cite{su3paper}: we set $\theta=0$ and the improvement coefficient 
$c_t(g_0)$ to its one-loop value in order to cancel most of the extra
${\rm O}(a)$ corrections which would be absent in the pure gauge theory
and are introduced through the Schr\"odinger functional
boundary conditions. 
In addition we used trivial 
background gauge fields. 

We start the discussion of our numerical results by showing in
fig.~{\ref{fig:corrfunction} a typical correlation function, 
$f_{2}(x_0)$ from eq.(\ref{corrfunc}), computed on a $16^4$ lattice.
The signal can be followed up to  
large distances and only for, say, $x_0\ge 3L/4$, the correlation
function becomes too noisy. A similar qualitative behaviour of our correlation functions 
was found
for other lattices, too. For the definition of $\bar{Z}$, we can use
$x_0=L/4$ or $x_0=L/2$, where the errors of the 
correlation functions are reasonably small. However, 
the continuum extrapolation of the results at $x_0=L/2$ appears more affected
by lattice artefacts, as already observed in ref.~\cite{roberto_pert}; therefore, 
in the following, we will
 only present results for the case $x_0=L/4$, i.e. we will
choose
\begin{equation}
\eta=\frac{1}{4}
\end{equation}
from now on. We remark that the case $L/a=6$ for $x_0=L/4$ is obtained by
an interpolation from the points at $x_0=(L\pm 2a)/4$ that cancels leading lattice artefacts. 
\begin{figure}
\vspace{0.0cm}
\begin{center}
\psfig{file=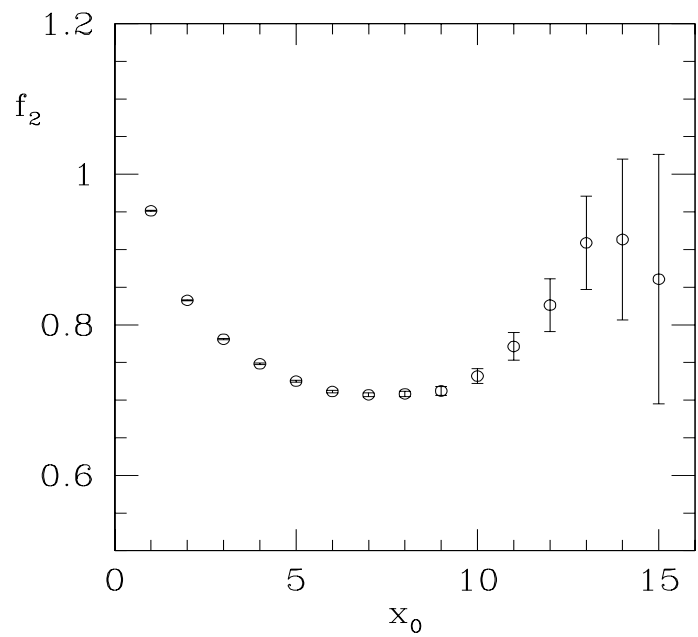, %
width=11cm,height=10cm}
\end{center}
\caption{ \label{fig:corrfunction} An example for a correlation function 
$f_{2}(x_0)$ in eq.~(\ref{corrfunc}), for a $16^4$
lattice. Parameters are $\beta\, = 7.0197$ and $\kappa\, = 0.144987$. 
}
\end{figure}
%
%

In order to obtain the running of the $Z(\mu)$, eq.~(\ref{finalZ}), 
we computed the step scaling functions for $\bar{Z}(L)$:
\begin{equation} \label{sigmabarz}
\sigma_{\bar{Z}} = \frac{\bar{Z}(2L)}{\bar{Z}(L)},
\end{equation}
for $f_1$
\begin{equation}\label{sigmaf1}
\sigma_{f_1} = \frac{\sqrt{f_1}(2L)}{\sqrt{f_1}(L)},
\end{equation}
and for the renormalization constant itself, 
\begin{equation}\label{sigmaz}
\sigma_{Z} = \frac{Z(2L)}{Z(L)}\;.
\end{equation}

We first discuss
$\sigma_{f_1}$ and $\sigma_{\bar{Z}}$. 
We have calculated $\sigma_{f_1}$ and $\sigma_{\bar{Z}}$
for five values of the coupling in 
the Schr\"odinger functional scheme renormalized at scale $L$. 
All results are reported in table 1. There we also give the values of
$\beta$ taken for the continuum extrapolation of the above-mentioned ratios. 
We recall that the values of $\beta$ are chosen such that when varying the number of lattice 
points $N$, the physical length $L$ of our box stays constant. 

\begin{table}[htbp]
  \begin{center}
    \leavevmode
    \begin{tabular}[]{|c|c|c|c|}
\hline
 $\beta$ & $a/L$  &  $\sigma_{\bar{Z}}$ & $\sigma_{f_1}$  \\
\hline\hline
 8.2415  & 0.0625 & 0.9270(14) & 0.9365(34) \\
 7.9993  & 0.0833 & 0.9282(13) & 0.9529(33) \\
 7.6547  & 0.1250 & 0.9244(14) & 0.9655(36) \\
 7.4082  & 0.1667 & 0.9162(15) & 0.9961(38) \\
\hline
 7.9560  & 0.0625 & 0.9181(18) & 0.9289(46) \\
 7.6985  & 0.0833 & 0.9223(15) & 0.9466(36) \\
 7.3632  & 0.1250 & 0.9194(15) & 0.9662(38) \\
 7.1214  & 0.1667 & 0.9060(17) & 0.9934(47) \\
\hline
 7.6101  & 0.0625 & 0.9104(17) & 0.9253(42) \\
 7.3551  & 0.0833 & 0.9090(28) & 0.9174(65) \\
 7.0197  & 0.1250 & 0.9079(17) & 0.9585(42) \\
 6.7807  & 0.1667 & 0.8931(20) & 0.9982(53) \\
\hline
 7.3686  & 0.0625 & 0.8981(21) & 0.9095(44) \\
 7.1190  & 0.0833 & 0.9041(18) & 0.9285(45) \\
 6.7860  & 0.1250 & 0.9020(28) & 0.9541(63) \\
 6.5512  & 0.1667 & 0.8824(27) & 1.0067(77) \\
\hline
 7.0203  & 0.0625 & 0.8762(24) & 0.9012(55) \\
 6.7750  & 0.0833 & 0.8867(24) & 0.9172(55) \\
 6.4527  & 0.1250 & 0.8792(21) & 0.9622(55) \\
 6.2204  & 0.1667 & 0.8729(22) & 1.0451(68) \\
\hline
    \end{tabular}
    \caption{The lattice step scaling functions used for the continuum
            extrapolations given in  table~2 and table~3.}
    \label{all_extra_table}
  \end{center}
\end{table}

In fig.~\ref{fig:stepall} we show our data for $\sigma_{\bar{Z}}$. The different figures 
are labeled by the value of the running coupling constant $\bar{g}^2(L)$, taken
at scale $L$, in the Schr\"odinger functional scheme. For $\sigma_{\bar{Z}}$ we see a 
marked curvature of the data when plotted as a function of $a/L$. We tried to fit the
data using a linear and quadratic (in $a/L$) ansatz. 
For the linear fit, we used the three data points with smallest values of $a/L$, whereas
for the quadratic fit we used all data points. 
For most cases, the value of
$\chi^2$ per degree of freedom (d.o.f.) was much better for the fit using a quadratic 
form. It seems that the use of non-improved Wilson fermions induces large
lattice artefacts that appear linear {\em and} quadratic in $a/L$. 
We think that these large lattice artefacts are due to the non-zero momentum
that we are using for computing our observable. This is in accord with the
remarks in \cite{roberto_pert}. 
As can be seen
in table~2, the values of $\sigma_{\bar{Z}}$, extrapolated to the
continuum, suffer from the large error induced by the use of the quadratic extrapolation.

\begin{figure}
\vspace{0.0cm}
\begin{center}
\psfig{file=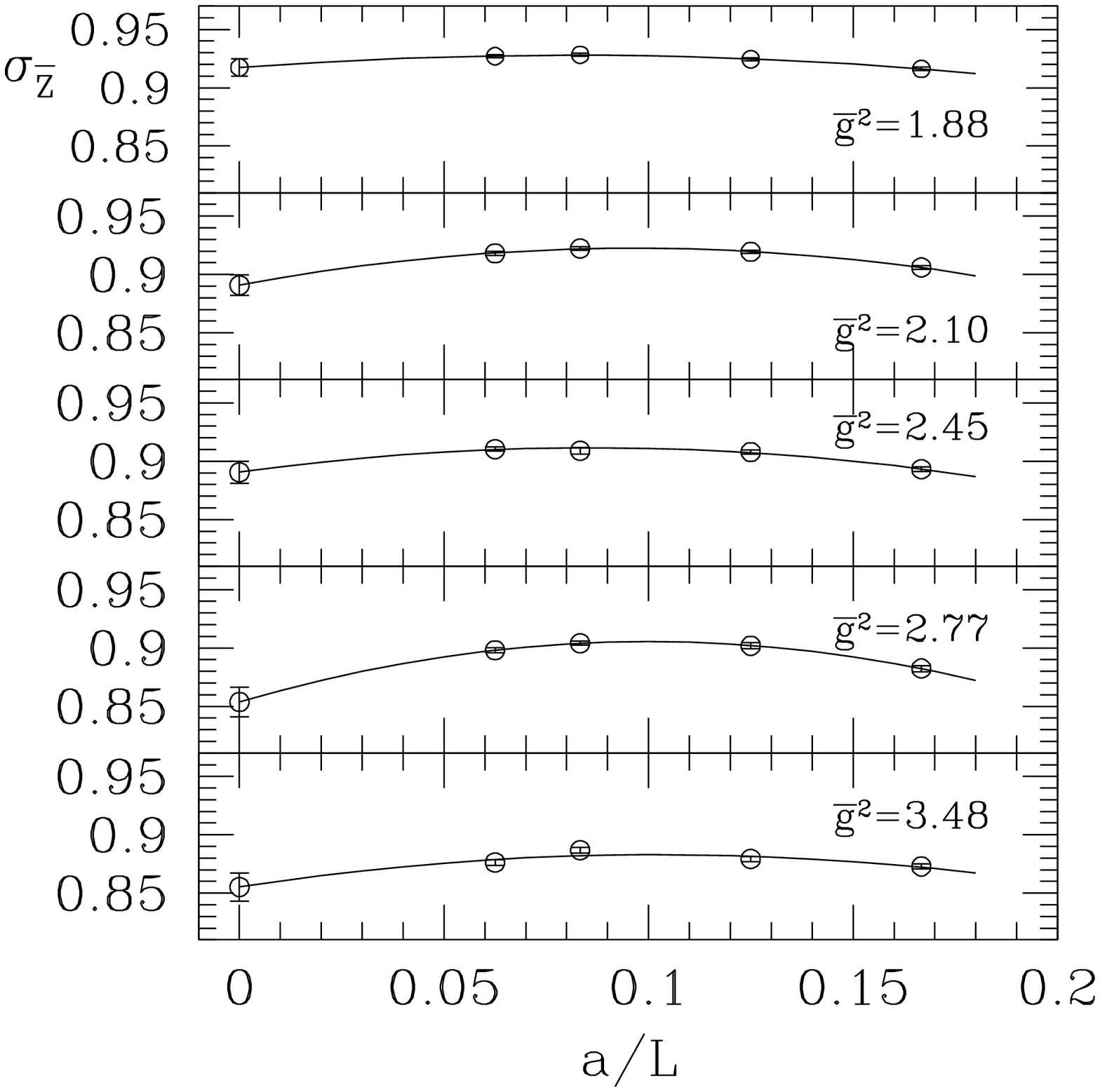, %
width=13cm,height=16cm}
\end{center}
\caption{ \label{fig:stepall} Continuum extrapolation of
$\sigma_{\bar{Z}}$ to obtain the
step scaling function for all values of the renormalized coupling considered. 
The solid line is 
a quadratic fit to the data. 
The values of $\bar{g}^2$,
labelling the graphs, are taken at scale $L$. 
}
\end{figure}

The situation is somewhat different for the continuum extrapolation 
of $\sigma_{f_1}$, shown in fig.~(3). Here we find that a linear 
fit in $a/L$, again leaving out the value of $\sigma_{f_1}$ 
for the  scaling from $L/a=6$ up to $L/a=12$, always gives a better or compatible $\chi^2/{\rm d.o.f.}$
as compared to a quadratic fit taking all data points. 
Since $f_1$ is computed at zero momentum,
the different behaviour of $\sigma_{f_1}$ as compared to 
$\sigma_{\bar{Z}}$ can be traced back again to the
lattice artefacts associated to the non-zero momentum present only in the unnormalized
constant $\bar{Z}$. 

\begin{figure}
\vspace{0.0cm}
\begin{center}
\psfig{file=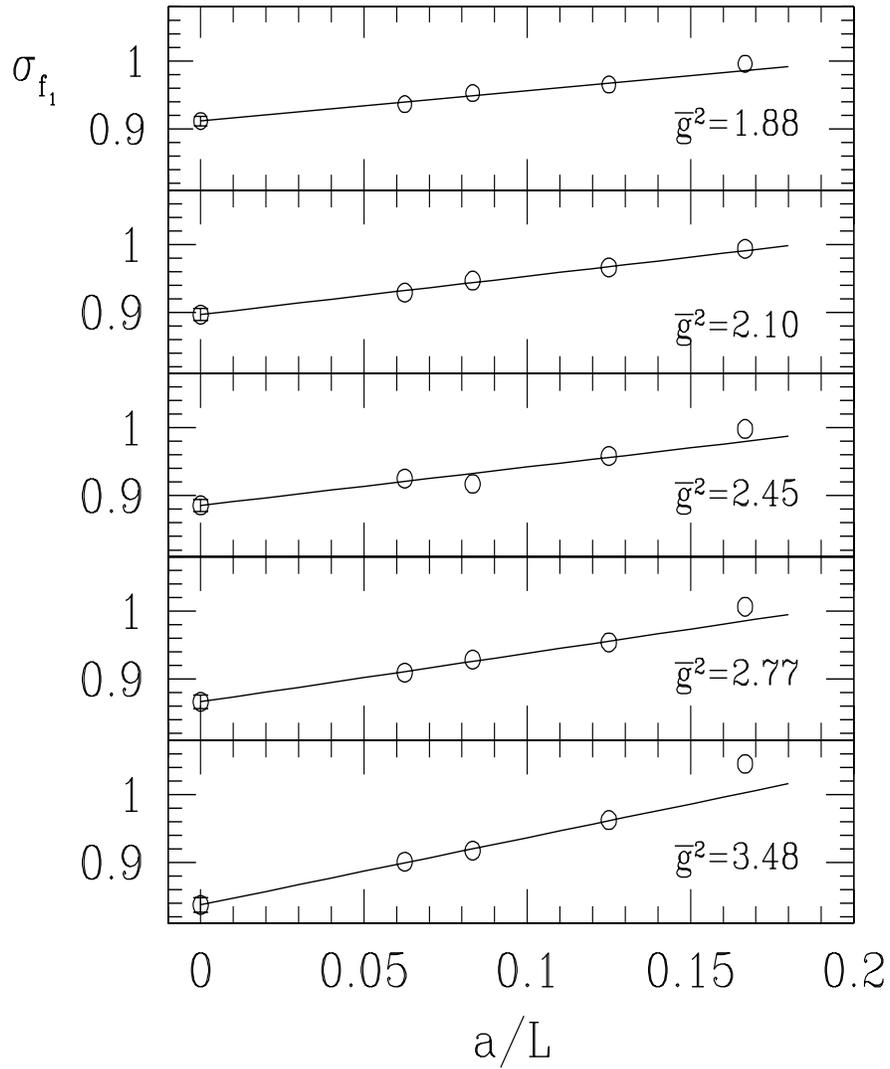, %
width=13cm,height=16cm}
\end{center}
\caption{ \label{fig:allf1} The continuum extrapolation
of
$\sigma_{f_1}$. The solid line is a linear fit to the data.
The values of $\bar{g}^2$,
labelling the graphs, are taken at scale $L$. 
The error bars in this case are smaller than the symbol sizes. 
}
\end{figure}


From the above discussion it follows that 
$\sigma_{f_1}$ and 
$\sigma_{\bar{Z}}$ 
have
to be extrapolated differently to the continuum limit, and we decided to 
perform this extrapolation independently for the step scaling functions. 
Such a strategy is certainly justified, given the fact that we did not find
any autocorrelation time in our data for $f_1$ and $\bar{Z}$, and that there
was only little correlations between the two quantities. (Remember that $f_1$
is computed at zero momentum, while $\bar{Z}$ needs a non-vanishing momentum.) 
We then finally compute the step scaling function for the 
renormalization constant of the
twist-2 non-singlet operator as the ratio of the individual continuum extrapolations of 
$\sigma_{\bar{Z}}$ and $\sigma_{f_1}$.

We summarize our results for all values of the different step scaling
functions in table~2, together with the corresponding values
of $\bar{g}^2(L)$ in the Schr\"odinger functional scheme. 
We summarize the continuum values of $\sigma_{Z}$ in table~3.
\begin{table}[htbp]
  \begin{center}
    \leavevmode
    \begin{tabular}[]{|c|c|c|c|c|c|}
\hline
$\bar{g}^2(L)$&$\sigma_{f_1}$&$\chi^2$&$\sigma_{\bar{Z}}$&$\chi^2$&Type of fit \\
\hline\hline
 1.8811       & 0.919(19)    & 3.95   & 0.917(8)   & 0.19   & quadratic \\
 2.1000       & 0.894(23)    & 1.26   & 0.891(9)   & 0.20   & quadratic \\
 2.4484       & 0.931(23)    & 4.24   & 0.891(9)   & 1.01   & quadratic \\
 2.7700       & 0.888(29)    & 1.66   & 0.854(13)  & 0.01   & quadratic \\
 3.48         & 0.913(30)    & 0.31   & 0.855(12)  & 6.44   & quadratic \\
\hline
 1.8811       & 0.912(7)     & 2.61   & 0.931(3)   & 1.53   & linear \\
 2.1000       & 0.897(9)     & 1.17   & 0.920(4)   & 3.60   & linear \\
 2.4484       & 0.886(9)     & 6.90   & 0.913(4)   & 0.03   & linear \\
 2.7700       & 0.866(10)     & 0.52   & 0.897(5)   & 3.64   & linear \\
 3.48         & 0.837(11)     & 0.39   & 0.880(5)   & 10.2   & linear \\
\hline          
    \end{tabular}
    \caption{The values of the step scaling functions are given {\em in the continuum}.
             The running coupling is computed in the SF scheme.}
    \label{table_fit_1}
  \end{center}
\end{table}

\begin{table}[htbp]
  \begin{center}
    \leavevmode
    \begin{tabular}[]{|c|c|}
\hline
 $\bar{g}^2(L)$ & $\sigma_{Z}$ \\
\hline\hline
 1.8811 & 1.006(12)  \\
 2.1000 & 0.993(15) \\
 2.4484 & 1.006(16) \\
 2.7700 & 0.985(20) \\
 3.48   & 1.021(21) \\
\hline          
    \end{tabular}
    \caption{The continuum values of the step scaling function for the renormalization
             constant of the twist-2 non-singlet parton density.}
    \label{table_fit_2}
  \end{center}
\end{table}

\section{The non-perturbative anomalous dimensions}

The numerical results discussed in the previous section allow us to 
extract the non-perturbative values of the anomalous dimensions.
We want to emphasize that, from now on, we are discussing results
that are already extrapolated to the continuum in the, maybe, somewhat unusual
Schr\"odinger functional scheme (instead of, say, the $\overline{\rm MS}$ scheme). 

The extraction of the anomalous dimension through the step scaling function,
in general requires the knowledge of
the beta function to the appropriate accuracy. For the running coupling $\bar{g}^2$
in the ${\rm SF}$ scheme, the beta function is well expressed by a three-loop formula 
up to $\bar{g}^2 = 3.5$ \cite{su3paper}.
However, 
the range of renormalized couplings explored in our work (see table~2 and in particular
the last two values $\bar{g}^2(L) = 2.77$ and  $\bar{g}^2(L) = 3.48$) lead to values for
$\bar{g}^2(2L)$ outside the domain of validity of the three-loop parametrization.
Therefore, the extraction from eq.~(\ref{sigma_pert}), valid to three-loop only, of  
the two- and three-loop coefficients of the perturbative
expansion of the anomalous dimensions in $\bar{g}^2(L)$, can only be done safely for
the first three values of $\bar{g}^2(L)$ simulated. 

To describe our data we have performed fits to the step scaling functions using
eq.~(\ref{sigma_pert}), i.e. by  expanding $\gamma(g^2)$ and $\beta(g^2)$ up to three-loop order,
eqs.~(\ref{gamma_3loop}), (\ref{beta_3loop}). 
For the fit for $\sigma_Z$ we have fixed the one- and two-loop contribution by
setting $\gamma_0$ and $\gamma_1$ to their perturbative values. 
The latter are extracted from a conversion from the modified minimal subtraction 
scheme results \cite{two_loops} to the SF scheme. It   
takes into account both the different operator renormalization constant $B_0$ in 
eq.~(\ref{Z_pert}):

\begin{eqnarray}
B_0 & = & 0.2635\;\; ({\rm SF}) \nonumber \\
B_0 & = & 0.0108\;\; (\overline{\rm MS})
\end{eqnarray}
\noindent and the relation between $g^2_{\overline{\rm MS}}$ and $g^2_{\rm SF}$ 
\cite{su3paper} 
\begin{equation}
g^2_{\overline{{\rm MS}}} = g^2_{\rm SF} + \frac{1.2556}{(4 \pi)}g^4_{\rm SF}.
\end{equation}

The perturbative two-loop coefficient to be expected is:
\begin{equation}
\gamma_1^{\rm SF} = -0.0268  \;\; [\mu^{-1}=L]\; ,
\end{equation}
\noindent where $\mu^{-1}=L$ indicates the scale at which the SF coupling is
renormalized, to be compared with
\begin{equation}
\gamma_1^{\overline{{\rm MS}}} = 0.0039. 
\end{equation}
The fit to our data of $\sigma_Z$ is shown in fig.~\ref{fig:sigma_Z_L}. 
We find for the three-loop coefficient $\gamma_2$:                                  
\begin{equation}
\gamma_2^{\rm SF} = 0.0034(7)\;\; [\mu^{-1}=L]\; .
\end{equation}
\noindent We also plot in fig.~\ref{fig:sigma_Z_L} the analytical form of the step scaling
function when we truncate $\gamma(g^2)$ to one- and two-loop only, 
while always keeping the three-loop expansion of $\beta(g^2)$. 
We remind that the fit is based on the first three points
only, although we show in fig.~\ref{fig:sigma_Z_L} also the remaining data points.
 
\begin{figure}
\vspace{0.0cm}
\begin{center}
\psfig{file=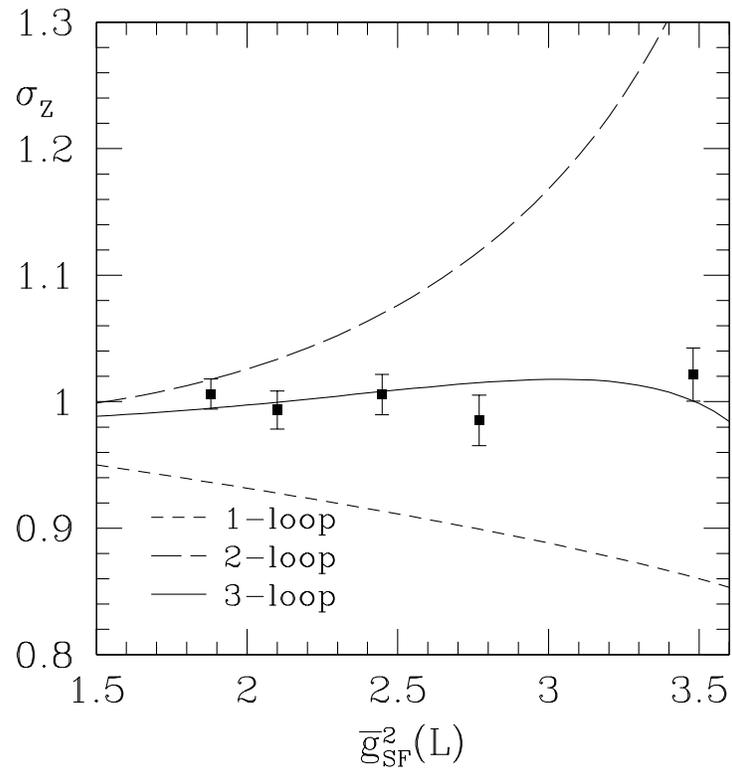, %
width=12cm,height=12cm}
\end{center}
\caption{ \label{fig:sigma_Z_L} The dependence of $\sigma_Z$ 
on the running coupling $\bar{g}^2_{\rm SF}(L)$ in the
Schr\"odinger functional scheme taken as $\eta=1$, i.e. the scale
to be $L$. 
}
\end{figure}

The corrections due to higher loops are large and oscillating in sign, which might be a signal of an
unfortunate choice of the  expansion parameter renormalized at the scale
$\mu^{-1}=L$.
Therefore, we have repeated the fit for the step scaling function by using a different scale for the
running coupling, $\bar{g}^2_{\rm SF}(L/4)$ instead of $\bar{g}^2_{\rm SF}(L)$. 
This opens the possibility of including 
in the fit the largest  $\bar{g}^2_{\rm SF}(L)$ points that correspond in the case of 
$\bar{g}^2_{\rm SF}(L/4)$ to still moderate
values where the three-loop parametrization is valid. The choice of the scale 
was motivated by the identification  with the value of $x_0=L/4$ where the operator is inserted and  
we did not attempt to optimize such a scale.

We show a fit to our data, employing the scale $L/4$ in the upper graph 
of fig.~\ref{fig:sigma_Z_f1_L4}. 
As above, the $\gamma_0$ and $\gamma_1$ were fixed to their perturbative values.
The new value for $\gamma_1^{\rm SF}$ obtained from eq.(39) after the change of scale is:
\begin{equation}
\gamma_1^{\rm SF} = -0.0181\;\; [\mu^{-1}=L/4]\; ,
\end{equation}
\noindent and we obtain from our fit for the three-loop coefficient:
\begin{equation}
\gamma_2^{\rm SF} = -0.005(3) \;\; [\mu^{-1}=L/4]\; ,
\end{equation}

\noindent with a $\chi^2$ per degree of freedom of $0.7$.
Figure~\ref{fig:sigma_Z_f1_L4} 
shows that, with the  choice of a smaller scale and of 
correspondingly smaller values for the running couplings, the relevance of higher-loop terms
decreases and a three-loop expression appears to be safe.

\begin{figure}
\vspace{0.0cm}
\begin{center}
\psfig{file=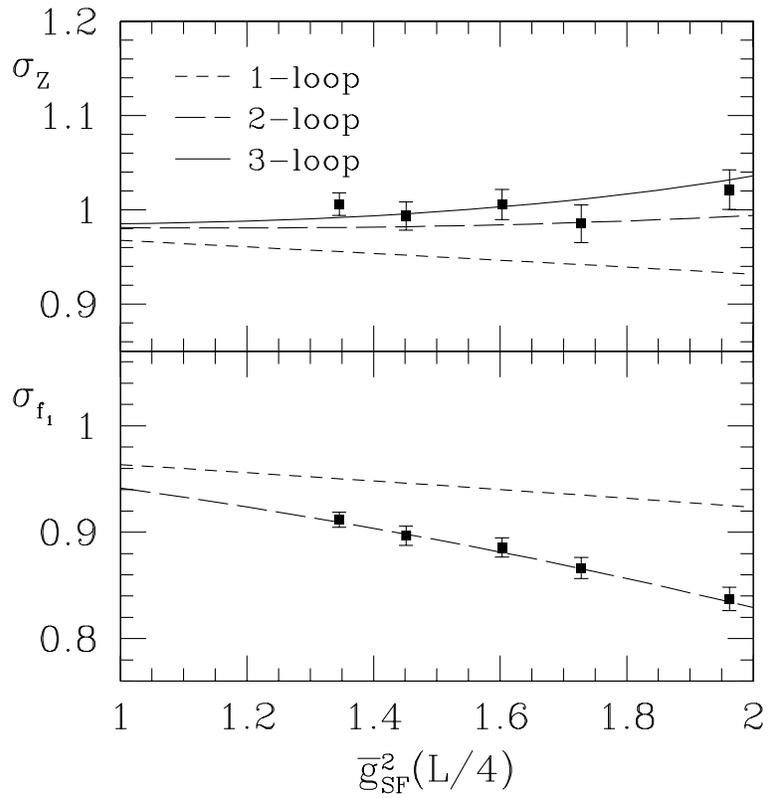, %
width=12cm,height=12cm}
\end{center}
\caption{ \label{fig:sigma_Z_f1_L4} The dependence of $\sigma_Z$ and $\sigma_{f_1}$ 
on the running coupling in the 
Schr\"odinger functional scheme evaluated at scale $L/4$, 
$\bar{g}^2_{\rm SF}(L/4)$. 
}
\end{figure}
 


We have also made a fit to the anomalous dimensions of the wave function renormalization constant 
$\sqrt f_1$:
in this case we can only fix $\gamma_0$ to its perturbative value $\gamma_0 =  0.05066$.
Fitting both the two- and three-loop coefficients leads to large errors in the fitted values.
We therefore attempted only a fit of the two-loop coefficient which gave a
reasonable value of $\chi^2=0.1$. Using the coupling $g^2_{\rm SF}(L/4)$, 
the fit turns out to be very stable when three, four
or five points are included in the fit. The value for the two-loop coefficient
obtained is
\begin{equation}
\gamma_1^{\sqrt{f_1}} = 0.030(3)\; ,
\end{equation}

%
 
The result for $g^2_{\rm SF}(L/4)$ is plotted in fig.~\ref{fig:sigma_Z_f1_L4} (lower graph).
In general, the sizeable two- and three-loop contributions for $Z$ and $f_1$ are mainly due
to the large constants $B_0$ appearing in the definition of the renormalization constants.
When such a constant is small, as in the case of the unnormalized $\bar{Z}$, the two-loop coefficient
is rather small, of the order of the $\overline{\rm MS}$ coefficient. We attempted a fit
to $\sigma_{\bar{Z}}$ and find:
\begin{equation}
\gamma_1^{\bar{Z}} = 0.003(3)\; ,
\end{equation}
\noindent indicating a dominant one-loop running.

\section{Conclusions}

We have studied the non-perturbative running of the average momentum of 
non-singlet parton densities in the SF scheme
in the region of renormalized $\alpha_{\rm SF}$ ranging from $0.1$ to $0.2$. 
From the step scaling functions we have extracted a three-loop
parametrization of the anomalous dimensions. In turn, from the knowledge of the three-loop 
anomalous dimensions, we can calculate the running of the renormalization constant $Z$. 
As a check, this running is compared in fig.~\ref{fig:running}  
with our results for the step scaling functions. 
The running in the SF scheme appears to be rather slow for the energy ranges that we have explored.
The errors of the data obtained in our numerical simulations are 
sizeable.
\begin{figure}
\vspace{0.0cm}
\begin{center}
\psfig{file=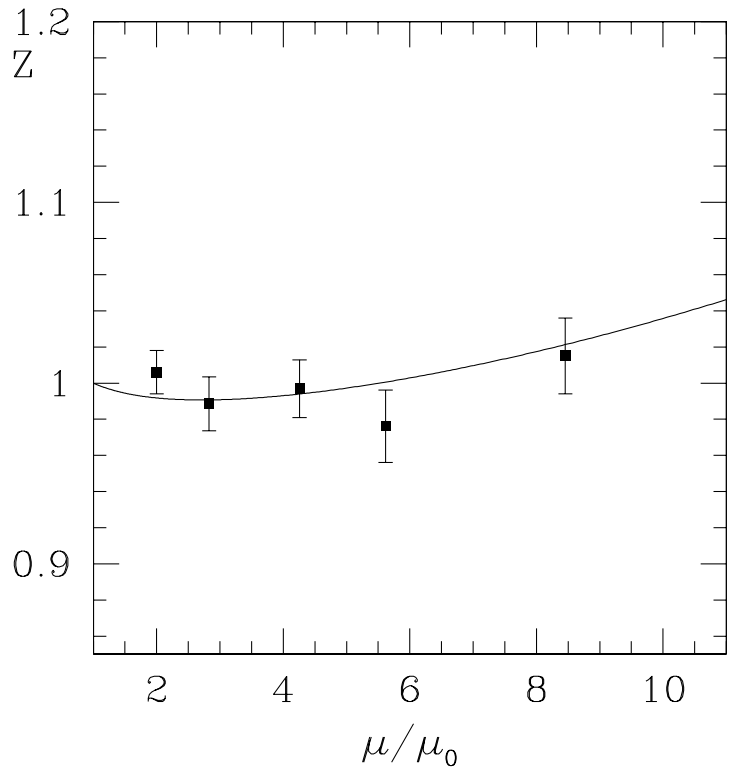, %
width=12cm,height=12cm}
\end{center}
\caption{ \label{fig:running} The dependence of $Z$ 
on the scale $\mu$ as compared to the three-loop running, where $\mu_0$ is the scale at
which $Z$ is normalized to unity. 
}
\end{figure}

The values of $\alpha_s$ covered in this work, 
according to ref.~\cite{su3paper}, correspond to the
energy range from $1$ to $10$ GeV.
Higher values of the renormalized couplings can be explored, and a higher accuracy for the continuum
extrapolation reached if leading lattice artefacts, mainly related to the lattice momentum
quantization, are removed, for example by using the non-perturbatively improved clover action
and, possibly, improved operators.
This calculation allows us to connect the continuum limit of a 
lattice evaluation of the hadron matrix element of the average momentum of non-singlet
parton densities, renormalized at a fixed
low-energy scale, with high energy experimental results, without an a priori unknown systematic
error deriving from the truncation of the perturbative series for the anomalous dimensions.

ACKNOWLEDGEMENTS. We thank H. Wittig for sending us,prior to publication, results for the values of $\beta$
used in this work to perform the matching. 
We are most grateful to M. L\"uscher, R. Sommer and P. Weisz
for critical discussions.


\begin{thebibliography}{99}
%
\bibitem{phenomeno} J.~Huston et al., 
                    Phys.Rev. D58 (1998) 114034.
%
\bibitem{martinelli}
G.~Martinelli, C.~Pittori, C.T.~Sachrajda, M.~Testa and A.~Vladikas,
Nucl. Phys. B445 (1995) 81.
%
\bibitem{schierholz1}
M. G\"ockeler, R.~Horsley, E.-M.~Ilgenfritz, H.~Perlt, P.~Rakow, G.~Schierholz and A.~Schiller,
Nucl. Phys. B472 (1996) 309.
%
\bibitem{schierholz2}
C.~Best, M. G\"ockeler, R.~Horsley, E.-M.~Ilgenfritz, H.~Perlt, P.~Rakow, A. Sch\"afer, G.~Schierholz, 
       A.~Schiller and S.~Schramm, Phys. Rev. D56 (1997) 2743.
%
\bibitem{schierholz3}
M.~G\"ockeler, R.~Horsley, H.~Oelrich, H.~Perlt, D.~Petters, P.E.L.~Rakow, A.~Sch\"afer, G.~Schierholz and 
   A.~Schiller,
preprint DESY-98097, hep-lat/9807044.
%
\bibitem{schrfunc} M.~L\"uscher, R.~Narayanan, P.~Weisz and U.~Wolff,
                    Nucl. Phys. B384 (1992) 168.
%
\bibitem{sint}
S.~Sint, Nucl. Phys. B421 (1994) 135.
%
\bibitem{paper3} M.~L\"{u}scher, S.~Sint, R.~Sommer, P.~Weisz and U.~Wolff,
                 Nucl. Phys. B491 (1997) 232.
%
\bibitem{letter} K.~Jansen, C.~Liu, M.~L\"{u}scher, H.~Simma, S.~Sint, R.~Sommer,
                 P.~Weisz and U.~Wolff, Phys. Lett. B372 (1996) 275.
%
\bibitem{quarkmass} S.~Capitani, M.~Guagnelli, M.~L\"uscher, 
                           S.~Sint, R.~Sommer, P.~Weisz and H.~Wittig, Nucl. Phys. B (Proc.Suppl.) 63 (1998) 153. 
%
\bibitem{roberto_pert} A.~Bucarelli, F.~Palombi, R.~Petronzio and A.~Shindler, CERN preprint, CERN/TH-98-251,
                       hep-lat/98-08-005.
%
\bibitem{rainer_lecture} R.~Sommer, DESY preprint, DESY 97-207, hep-lat/9711243. 
%
\bibitem{martin_lecture} M.~L\"uscher, DESY preprint, DESY 98-017, hep-lat/9802029. 
%
\bibitem{paper1} M.~L\"{u}scher, S.~Sint, R.~Sommer and P.~Weisz,
                 Nucl. Phys. B478 (1996) 365.
%
\bibitem{paper4}  M.~L\"uscher, S.~Sint, R.~Sommer and H.~Wittig,
                  Nucl. Phys. B491 (1997) 344.
%
\bibitem{peterstefan} S.~Sint and P. Weisz, preprint, MPI-PhT/98-50, hep-lat/9808013.       
%
\bibitem{capitani} S.~Capitani and G.~Rossi, Nucl. Phys. B433 (1995) 351.
%
\bibitem{two_loops} E.G.~Floratos, D.A. Ross and C.T. Sachrajda,
                    Nucl. Phys. B129 (1977) 66. Erratum: B139 (1978) 545.
%
\bibitem{su3paper} M.~L\"uscher, R.~Sommer, P.~Weisz and U.~Wolff,
       Nucl. Phys. B413 (1994) 481.
%
\bibitem{ssor}
S. Fischer et. al., Comp. Phys. Comm. 98 (1996) 20;\
N. Eicker et. al., Nucl. Phys. B (Proc. Suppl.) 63 (1998) 955.
%
\bibitem{bicg} A.~Frommer, V.~Hannemann, T.~Lippert, B.~Noeckel and K.~Schilling,
               Int. J. Mod. Phys. C5 (1994) 1073.
%
\end{thebibliography}
\end{document}